\documentclass[%
 reprint,
 aip,
showkeys,
 amsmath,amssymb,
longbibliography,
]{revtex4-1}

\usepackage{graphicx}
\usepackage[caption=false]{subfig} 
\usepackage{pifont}
\usepackage{accents} 
     
\usepackage{dcolumn}
\usepackage{bm}

\newcommand{\ee}{\end{equation}} 
\newcommand{\be}{\begin{equation}}

\usepackage[mathscr]{euscript}  
\usepackage{xcolor}  

\usepackage{tcolorbox}
\usepackage{comment}
\usepackage{float}

\makeatletter
\newsavebox{\@brx}
\newcommand{\llangle}[1][]{\savebox{\@brx}{\(\m@th{#1\langle}\)}%
  \mathopen{\copy\@brx\kern-0.5\wd\@brx\usebox{\@brx}}}
\newcommand{\rrangle}[1][]{\savebox{\@brx}{\(\m@th{#1\rangle}\)}%
  \mathclose{\copy\@brx\kern-0.5\wd\@brx\usebox{\@brx}}}
\makeatother

\def\dbar{{\mkern3mu\mathchar'26\mkern-12mu d}}
\makeatletter
\newcommand{\vast}{\bBigg@{3.4}}
\makeatother

\usepackage{tikz}

\makeatletter
\newcommand\emailx[1]{%
\move@AF%
\def\@affil{{\normalfont\,#1\strut}{}}%
}%

\begin{document}

\preprint{ApS/123-QED}

\title{Enhanced Efficiency in Shear-Loaded Brownian Gyrators}

\author{Iman Abdoli}
\email{iman.abdoli@hhu.de}
\affiliation{Institut für Theoretische Physik II - Weiche Materie, Heinrich-Heine-Universität Düsseldorf, Universitätsstraße 1, D-40225 Düsseldorf, Germany}

\author{Abhinav Sharma}
\affiliation{Mathematisch-Naturwissenschaftlich-Technische Fakultät, Institut für Physik,
	Universität Augsburg, Universitätsstraße 1, D-86159 Augsburg, Germany}
\affiliation{Leibniz-Institut für Polymerforschung Dresden, Institut Theory der Polymere, Hohe Straße 6, D-01069 Dresden, Germany}

\author{Hartmut L\"{o}wen}
\affiliation{Institut für Theoretische Physik II - Weiche Materie, Heinrich-Heine-Universität Düsseldorf, Universitätsstraße 1, D-40225 Düsseldorf, Germany}

\begin{abstract} 
A Brownian gyrator is a system in which a particle experiences thermal noise from two distinct heat baths. This nonequilibrium setup inherently generates a nonzero torque, leading to gyrating motion around a potential energy minimum. As a minimal model for a heat engine, the Brownian gyrator provides valuable insights into energy conversion and nonequilibrium dynamics. 
Here, we investigate the effect of an externally imposed shear flow on a Brownian gyrator, treating it as a mechanical load. The shear flow introduces a tunable mechanism that allows the system to operate either as a heat engine, extracting work from the temperature gradient, or as a refrigerator, transferring heat from the colder to the hotter bath. Focusing on the heat engine regime, we analytically derive the steady-state probability distribution to compute the average torque exerted by the gyrator and quantify the mechanical power extracted from the shear. Our results show a remarkable increase in efficiency compared to the standard Brownian gyrator without shear, approaching Carnot efficiency at maximum power. Surprisingly, we also find that while the system can operate efficiently as a heat engine, it may become unstable before reaching the stall condition, highlighting a fundamental trade-off between efficiency and stability in shear-driven microscopic engines.

\end{abstract}

\maketitle

\section{Introduction}

Thermal fluctuations drive microscopic systems out of equilibrium, giving rise to intriguing transport phenomena in stochastic thermodynamics.
Within this realm, the concept of the Brownian gyrator has gained attention as a minimal model for examining systems under non-equilibrium conditions\cite{filliger2007brownian, chiang2017electrical, bae2021inertial, dos2021stationary, movilla2021energy, squarcini2022fractional, cerasoli2022spectral}. 
Originally introduced to explore energy transfer in systems subject to thermal noise from two distinct heat baths, the Brownian gyrator consists of a particle diffusing in a two-dimensional anisotropic harmonic potential, with each spatial degree of freedom coupled to a heat bath at a different temperature~\cite{abdoli2022escape, miangolarra2024energy, miangolarra2024minimal, miangolarra2024stochastic}. This temperature asymmetry, combined with a misalignment between the potential’s principal axes and the temperature axes, generates steady-state rotational currents or gyration, which can be interpreted as an emergent torque~\cite{filliger2007brownian,pietzonka2018universal}. 
Crucially, the gyrating motion vanishes in the absence of spatial cross-correlations, which can originate either from the structure of the potential or from an external magnetic field~\cite{abdoli2020correlations, abdoli2022tunable, muhsin2025active}. Theoretical predictions of this model have been experimentally validated by applying a strongly fluctuating electric field along one direction, effectively simulating the role of an additional temperature~\cite{argun2017experimental}.

While Brownian gyrators illustrate how temperature asymmetry can drive rotational motion, shear flow introduces another mechanism for directed transport in non-equilibrium systems. Shear forces arise naturally in various contexts, from colloidal suspensions to biological fluids and microfluidic devices~\cite{bergenholtz2002non, lowen2005colloidal, ballesta2008slip, stone2004engineering}. In non-equilibrium statistical physics, shear modifies diffusion by inducing anisotropic transport and velocity correlations~\cite{brady1993brownian, fuchs2002theory, brader2008first}. In confined environments, such as optical traps or microfluidic channels, it introduces cross-correlations between spatial degrees of freedom, biasing particle trajectories and altering displacement statistics~\cite{howard2016axial, holzer2010dynamics, bammert2010probability}. Beyond classical Brownian motion, the interplay between thermal fluctuations and shear flow produces rich transport behaviors. In a quiescent medium, diffusive motion follows the Einstein-Smoluchowski description~\cite{einstein1906theorie, von1906kinetischen}, but shear introduces more complex effects such as Taylor-Aris dispersion, resonant transport, flow-enhanced mixing, and shear-induced particle migration in Poiseuille flow, where colloids drift toward low-shear regions~\cite{taylor1953dispersion, aris1956dispersion, belongia1997measurements, huang2011direct, kahlert2012resonant, kumar2021taylor, marmet2017shear, C5SM01693B, PhysRevE.95.012610, abdoli2024shear}. These modifications become even more pronounced in systems where the particles exhibit internal activity or deformability. Recent studies on active Brownian particles have revealed that self-propulsion and shape fluctuations, when coupled with shear forces, can give rise to novel steady states, nontrivial cross-correlations, and enhanced transport properties~\cite{ten2011brownian, tarama2013dynamics, sandoval2018self, asheichyk2021brownian}.

Experimental and theoretical studies have revealed that shear-driven dynamics can lead to anomalous diffusion and non-trivial steady states, providing deeper insights into tracer transport in flow fields~\cite{ziehl2009direct, orihara2011brownian}. Recent advances in microfluidics and optical tweezers have enabled precise control of shear forces at microscopic scales, allowing for systematic investigations of transport phenomena in confined systems~\cite{faucheux1994confined, squires2005microfluidics, whitesides2006origins}. By applying controlled shear to colloidal particles and biological systems, these techniques have shed light on the interplay between thermal noise, deterministic driving, and dissipation. Notably, confining particles in harmonic traps, such as optical tweezers, has proven particularly effective in probing emergent cross-correlations induced by shear flow~\cite{ziehl2009direct}. Unlike free particles, which experience unconstrained motion and larger fluctuations, confined systems exhibit well-defined steady states, making them ideal for studying stochastic transport under controlled flow conditions.
In this work, we introduce and analyze a shear-loaded Brownian gyrator, where a trapped particle subjected to anisotropic thermal noise is also exposed to an external shear force. The shear flow acts as a controllable load, modifying the natural gyration and influencing the energy conversion properties of the system. By tuning the shear direction and magnitude, we explore the rich dynamical behavior of this system, including regimes where the shear opposes the natural gyration, leading to energy extraction. In contrast to macroscopic engines with distinct compression and expansion phases, the Brownian gyrator engine functions continuously, where heat absorption, work extraction, and dissipation occur simultaneously. The system sustains a steady-state cycle where it absorbs energy from the hot bath, converts a portion into work against the shear load, and dissipates the remaining heat into the cold bath. This continuous cycling highlights the unique nature of stochastic heat engines, where power extraction and efficiency emerge from a dynamic interplay of fluctuating forces rather than discrete thermodynamic steps~\cite{PhysRevLett.122.140601}.

Our analysis reveals that the interplay between shear flow and the intrinsic gyration of a Brownian particle gives rise to a highly tunable microscopic heat engine with strikingly rich behavior. By treating shear as an external load, we identify conditions where the system extracts mechanical power from thermal fluctuations and operates as a heat engine, while in other regimes, it functions as a refrigerator, transferring heat against the natural temperature gradient. Crucially, we find that the stalling shear rate—the point at which the gyration is fully suppressed—depends sensitively on the shear direction relative to the temperature axes. Surprisingly, a distinct parameter region emerges where the system remains an efficient engine but loses stability before reaching stall conditions, highlighting an intriguing trade-off between efficiency and stability.

A deeper investigation into the system's performance under shear reveals that for certain orientations, the extracted mechanical power reaches a well-defined maximum before vanishing at the stalling condition. Remarkably, the efficiency at maximum power can approach the Carnot limit, an uncommon and highly sought-after feature in finite-time thermodynamics. This enhancement occurs when the shear opposes the natural gyration, effectively harnessing energy from the temperature gradient in an optimized manner. However, the system’s stability imposes a fundamental constraint: for some configurations, it becomes unstable before reaching stall conditions, suggesting that while high efficiency is achievable, it comes at the cost of increased fluctuations that may disrupt steady-state operation.

This instability can be traced back to the complex interaction between shear-induced forces and the potential landscape. Unlike an isolated Brownian gyrator, which remains stably confined, the introduction of shear modifies the steady-state probability distribution and induces a net radial force that can expand, contract, or destabilize the system. In some cases, shear flow effectively acts as an additional confining force, reinforcing stability, while in others, it amplifies fluctuations and leads to destabilization before the engine reaches its theoretical performance limits. These results highlight a fundamental trade-off in non-equilibrium thermodynamics: optimizing efficiency requires a careful balance between stability and dissipation, particularly when external forces are present.

In the following sections, we introduce the mathematical model of the shear-loaded Brownian gyrator, derive the key thermodynamic quantities of a heat engine, and systematically analyze its efficiency and stability. 


\section{Model}

\begin{figure}[t]
	\centering
	\includegraphics[width=8cm]{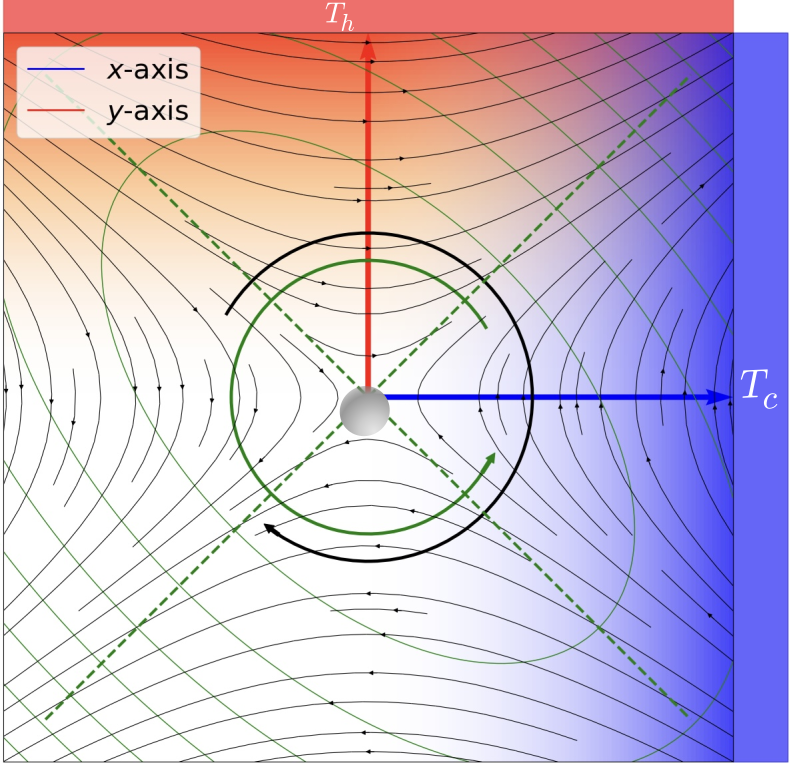}
    \caption{Schematic of a Brownian gyrator under shear flow. A Brownian particle (gray sphere) is trapped in an anisotropic harmonic potential and coupled to two heat baths at different temperatures, $T_h$ (hot) and $T_c$ (cold), leading to a non-equilibrium steady state with persistent gyration or torque (green curved arrow). The contours (green) represent the anisotropic potential, whose principal axes (dashed green lines) are misaligned with the temperature axes, which coincide with the coordinate axes. The shear flow, represented by black streamlines, acts as an external load on the system, opposing the natural gyration. The black curved arrow indicates the average torque induced by shear, which competes with the torque generated by the temperature gradient.}
	\label{figure01}
\end{figure}

We consider a Brownian particle confined in the anisotropic harmonic potential, $V(\mathbf{r})=\frac{1}{2}\mathbf{r}^\top\cdot\hat{U}\cdot\mathbf{r}$ subjected to the linear shear flow $\mathbf{F^{\text{shear}}} = \dot{\gamma}(y\cos(\theta), x\sin(\theta))$, at an angle $\theta$ with respect to the $x$-axis. The particle is connected to two thermal reservoirs at different temperatures $T_c$ (cold) and $T_h$ (hot), along the $x$ and $y$ axes, respectively. Here $\dot{\gamma}$ is the shear rate and $\hat{U} = k\begin{pmatrix}
	1 & \tilde{u} \\ \tilde{u} & 1 
\end{pmatrix}$ where $k$ and $u=k\tilde{u}$ are the stiffness of the potential and the potential coupling parameters, respectively. The probability for finding the particle at position $\mathbf{r}= (x, y)^\top$ at time $t$, $\rho(\mathbf{r}, t)$, can be described by the following Fokker-Planck equation 

\begin{equation}
	\label{FPE_time}
	\frac{\partial \rho(\mathbf{r}, t)}{\partial t} = -\nabla\cdot\left[\hat{\mathbf{A}}\mathbf{r} -  \hat{\mathbf{D}}\nabla \right]\rho(\mathbf{r}, t),
\end{equation}
where  $\hat{\mathbf{D}}=\frac{1}{\gamma}\mathrm{diag}(T_x, T_y)$ is the diffusion matrix and the drift matrix $\hat{\mathbf{A}}$ is given by

\begin{equation}
	\label{driftmatrix}
	\hat{\mathbf{A}} = -\frac{k}{\gamma}\begin{pmatrix}
		1 & \tilde{u}-\dot{\tilde{\gamma}}\cos(\theta) \\ \tilde{u}-\dot{\tilde{\gamma}}\sin(\theta) & 1 
	\end{pmatrix},
\end{equation}
where 
\begin{align}
\dot{\tilde{\gamma}} & = \dot{\gamma}/k, \\
\tilde{u} & = u/k,
\end{align}
are dimensionless parameters. This formulation explicitly captures the interplay between the potential-induced restoring forces and the external shear-driven dynamics. The presence of off-diagonal terms in $\hat{\mathbf{A}}$ highlights the coupling between the $x$ and $y$ coordinates due to both the potential coupling and the shear force. An illustration of the model is presented in Fig.~\ref{figure01}. 

Since equation~\eqref{FPE_time} is linear, the steady-state solution follows a multivariate Gaussian distribution which in the steady state reads (see appendix~\ref{appendixA})
\begin{figure}[t]
	\centering
	\includegraphics[width=8.6cm]{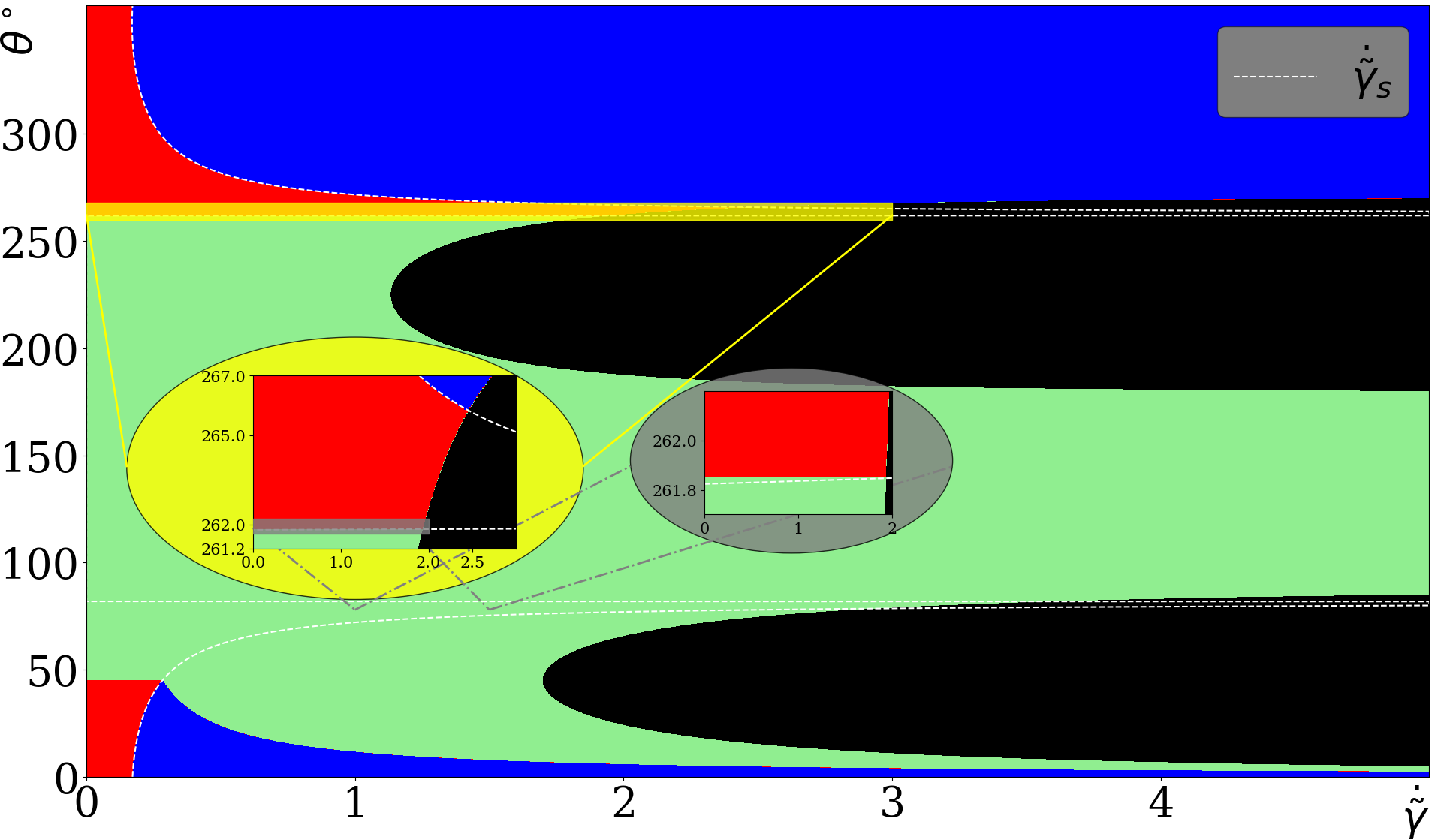}
	\caption{Phase diagram of the system's operational regimes as a function of the shear rate and shear direction. Red regions indicate where the system functions as a heat engine, extracting work from the thermal gradient and shear. 
	Blue regions correspond to the refrigerator regime, where the system absorbs heat from the colder bath and transfers it to the hotter bath.
	Green regions represent areas where the system is stable but does not function as an engine or refrigerator. Black regions mark instability, meaning the system cannot maintain steady-state operation. The white dashed line represents the stalling shear rate—the point at which the gyration is fully suppressed— beyond which the system ceases to function as a heat engine. The yellow zoomed-in region highlights an important instability phenomenon: the system can operate efficiently as a heat engine at intermediate values of shear rate, but it becomes unstable before reaching the stalling point. The gray zoomed-in region shows that the stalling shear rate does not  coincide with the interface between the green and red regions. 
	The parameters are chosen to allow the system to operate as an efficient heat engine, with $\beta = T_h/T_c = 7.0$ and $\tilde{u} = u/k = 0.2$. 
	}
	\label{figure02}
\end{figure}
\begin{equation}
	\label{pdf_ss}
	\rho(\mathbf{r}) = \frac{1}{Z}e^{-\mathbf{r}^\top\cdot\hat{\mathbf{C}}^{-1}\cdot\mathbf{r}},
\end{equation}
where $\hat{\mathbf{C}}^{-1}=\begin{pmatrix}
	\mu_1 & \mu_3 \\ \mu_3 & \mu_2 
\end{pmatrix}$ is the inverse of the steady-state covariance matrix, which determines key statistical properties of the system and whose elements are given as


\begin{align}
	\mu_1 &\! = \!  \frac{ 2\beta \!+\! \tilde{u}^{2} (1\! -\! \beta)\! +\! \dot{\tilde{\gamma}} \left[ S \left( \tilde{u} (\beta\! -\! 2)\! +\! \dot{\tilde{\gamma}} S \right)\! +\! C(\tilde{u}\! -\! \dot{\tilde{\gamma}} S) \beta \right] }{L^2\mathcal{D}}, \label{mu1}\\
	\mu_2 & \!=\!  \frac{ 2\! +\! \tilde{u}^{2} (\beta\! -\! 1)\! +\! \dot{\tilde{\gamma}} \left[ \beta \dot{\tilde{\gamma}} C^2\! +\! \tilde{u} S\! +\! C (\tilde{u} (1\! -\! 2\beta)\! -\! \dot{\tilde{\gamma}} S ) \right] }{L^2\mathcal{D}}, \label{mu2}\\
	\mu_3 & =  \frac{ \tilde{u} (1 + \beta) - \dot{\tilde{\gamma}} (S + \beta C) }{L^2\mathcal{D}}. \label{mu3}
\end{align}
The normalization factor \( Z \) is given by:
\begin{equation}
	Z = \frac{ \sqrt{1 - \tilde{u}^{2} + \dot{\tilde{\gamma}} (\tilde{u} S + C (\tilde{u} - \dot{\tilde{\gamma}} S))}}{\pi L^2 \sqrt{\mathcal{D}}}. \label{normalization}
\end{equation}
where $L=\sqrt{T_c/k}$ is the characteristic length,	$\beta = T_h/T_c$, 
$S = \sin(\theta)$,  $C = \cos(\theta)$ 
 and $	\mathcal{D} = \tilde{u}^{2} (1 - \beta)^2 + 4\beta + \dot{\tilde{\gamma}} (\beta C - S) \left(-\dot{\tilde{\gamma}} S + 2 \tilde{u} (1 - \beta) + \dot{\tilde{\gamma}} \beta C \right)$.

\begin{figure*}[t]
	\centering
	\includegraphics[width=\linewidth]{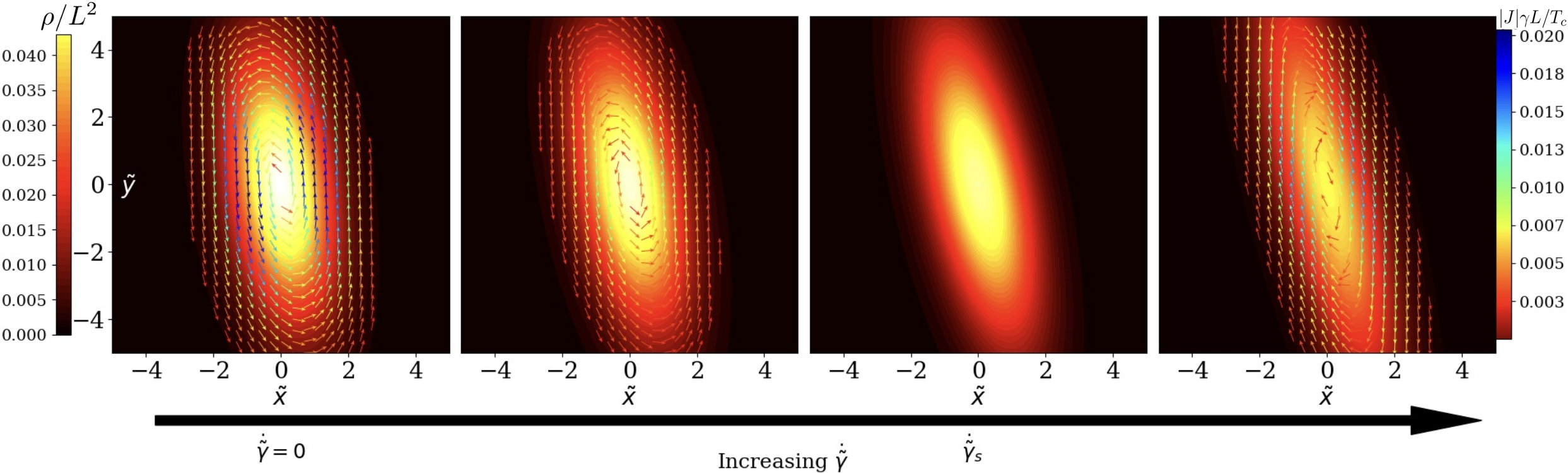}
	\caption{     Steady-state probability density \(\rho/L^2\) and corresponding probability fluxes for a Brownian gyrator under shear at \(\theta = 270^\circ\), \(\beta = 7.0\), and \(\tilde{u} = 0.2\), for increasing shear rate \(\dot{\tilde{\gamma}}\). Here \(\tilde{x}=x/L\) and \(\tilde{y}=y/L\) where \(L=\sqrt{T_c/k}\) is the characteristic length. The color scale  (left colorbar) represents the probability density, while the arrows depict the direction of the probability flux with color-coded magnitude. The first figure (leftmost) shows the system in the absence of shear, where the gyrator exhibits a counter-clockwise rotational motion due to the  anisotropic noise. As shear is introduced against this natural gyration, the rotational flux weakens, stalls at the stalling shear rate \(\dot{\tilde{\gamma}}_s\), and reverses beyond this point. The transition from a dominant temperature-driven gyration to a shear-dominated motion is clearly visible as \(\dot{\tilde{\gamma}}\) increases.}
	\label{figure03}
\end{figure*}
We examine the stability conditions required for the system to maintain steady-state gyration. The stability of the probability distribution is determined by the eigenvalues of the drift matrix $\hat{\mathbf{A}}$. For the system to reach a steady state, the real parts of these eigenvalues must be negative, ensuring that perturbations in the system decay over time rather than grow indefinitely. This stability criterion defines a parameter space within which the system remains well-behaved and confined, preventing unbounded trajectories or diverging probability fluxes. Out of this set of stable regimes, we identify specific conditions where the system extracts mechanical power from thermal fluctuations, thereby functioning as a microscopic heat engine. By treating the shear force as a mechanical load, we explore how its orientation and magnitude influence the performance of the engine, allowing us to pinpoint regions of high efficiency and power output.

In Fig. \ref{figure02}, we present the phase diagram of the system, identifying stable and unstable regions as a function of shear rate and direction. Stability is a prerequisite for maintaining steady-state gyration, and we focus on the parameter space where the system remains stable while extracting mechanical power. The green region marks conditions where the system is stable but does not function as a heat engine, whereas the black regions correspond to instability, where fluctuations grow uncontrollably. The red region highlights the set of stable parameters where the system extracts work against the shear force, operating as a heat engine. Notably, in some cases, the system becomes unstable before reaching the stalling shear rate (white dashed line), limiting the range of usable parameters for efficient energy conversion. 

Using the steady-state probability density in Eqs.~\eqref{pdf_ss}-\eqref{normalization}, we can calculate the probability fluxes in the system as

\begin{equation}
	\label{fluxes}
	\mathbf{J}(\mathbf{r}) = \left[\hat{\mathbf{A}}\mathbf{r} -  \hat{\mathbf{D}}\nabla \right]\rho(\mathbf{r}).
\end{equation}

Figure~\ref{figure03} represents the steady-state probability density and probability fluxes for the chosen operational regime, where the system functions as a heat engine. The probability density, shown in color, highlights how the particle distribution deforms under increasing shear. In the absence of shear flow, the natural gyration due to the temperature gradient dominates. As the shear rate increases, this rotation weakens, eventually stalling at a critical shear rate. Beyond this point, the direction of gyration reverses. The overlaid probability flux vectors illustrate this transition, showing how shear suppresses and ultimately inverts the circulation. This transition marks the boundary where the system ceases to operate as a heat engine, as no net work can be extracted against the shear load. The structural deformation of the steady-state distribution under shear is key to understanding the interplay between the temperature asymmetry and external driving forces in this system.

\section{Heat engine}
For the chosen operational regime, where the system functions as a heat engine, we now examine its performance by analyzing key thermodynamic quantities. We investigate the average torque exerted on the potential due to both temperature asymmetry and shear load, the mechanical power extracted, and the heat transfer from the hot bath. By evaluating these quantities, we determine the efficiency of the system and identify the conditions under which the Brownian gyrator achieves optimal performance. Finally, we explore how the gyrator can operate at maximum efficiency while simultaneously achieving maximum power.

\subsection{Average torque}
To extract mechanical power from the system, the applied shear is considered to act as a load, meaning that the torque induced by the shear flow opposes the natural gyration caused by the temperature asymmetry. This condition can be identified by analyzing the total average torque exerted on the potential. Specifically, the system must transition through a regime where the total average torque changes sign, ensuring that the contribution from shear counteracts the inherent rotational motion of the Brownian gyrator. This investigation provides insight into the parameter space where the system can be efficiently operated as a microscopic heat engine.


We utilize the steady-state probability density to compute the average torque, \(\langle\tau\rangle=\int\rho(\mathbf{r})(\mathbf{r}\times\mathbf{F})d\mathbf{r}\),  exerted by the particle on the potential.
Here \(\mathbf{F}=\mathbf{F}^{\text{shear}}+\mathbf{F}^{\text{c}}\) where $\mathbf{F}^c(\mathbf{r})=-\nabla V(\mathbf{r})$ is the conservative force. The average torque is given by
\begin{equation}
	\label{average_torque}
	\langle \tau\rangle=\left[\tilde{u}(T_h-T_c)-\dot{\tilde{\gamma}}(T_h\cos(\theta)-T_c\sin(\theta))\right],
\end{equation}
which consists of two distinct contributions. The first term originates from the temperature gradient, modulated by the potential coupling parameter $\tilde{u}$, which accounts for the misalignment between the principal axes of the harmonic potential and the temperature anisotropy. The second term arises from the applied shear flow, whose influence depends on both its magnitude and orientation, affecting the net torque on the system.

\begin{figure}
	\centering
	\includegraphics[width=8cm]{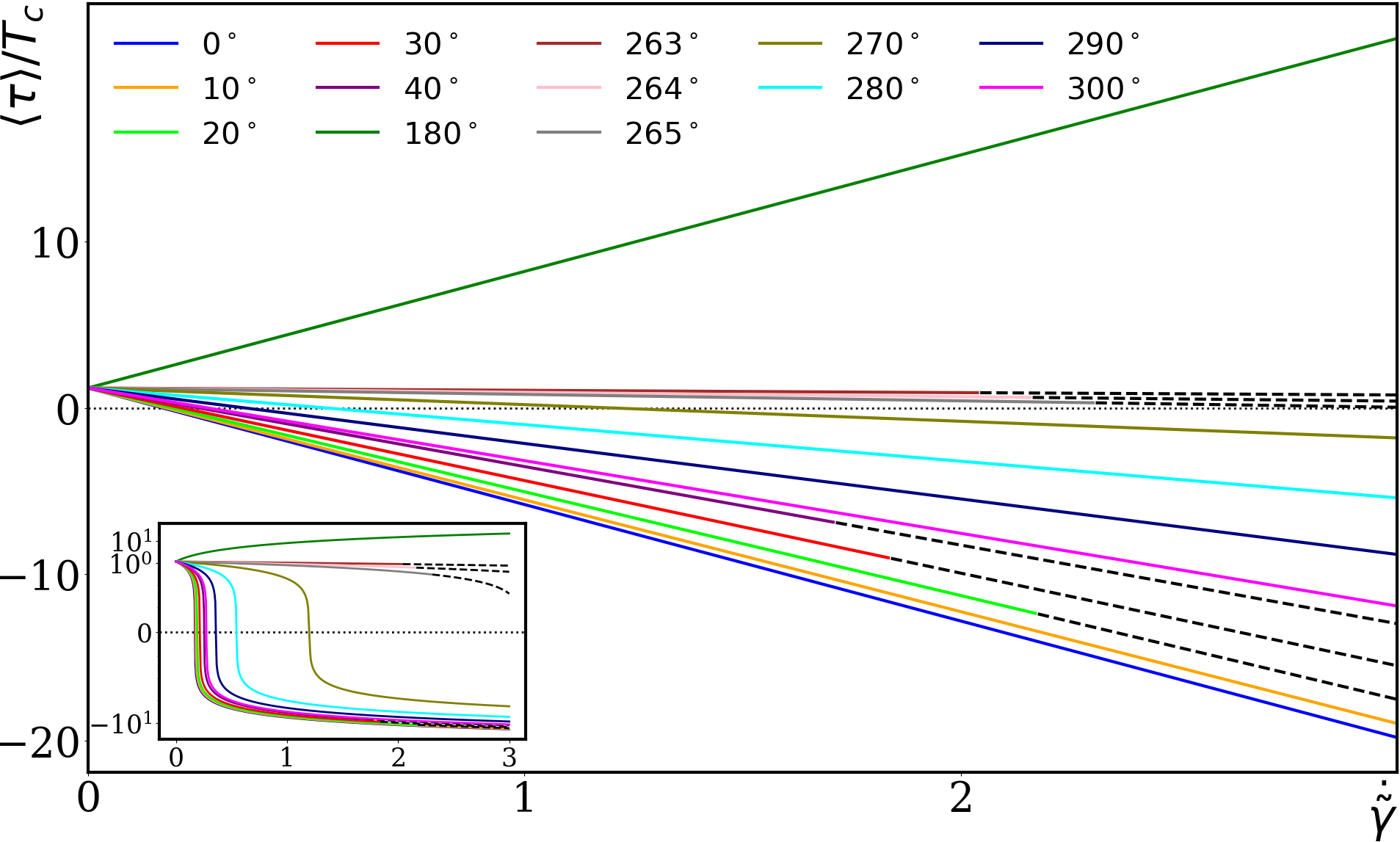}
	\caption{     Average exerted torque \(\langle \tau \rangle/T_c\) on the potential as a function of the shear rate \(\dot{\tilde{\gamma}}\) for different shear directions \(\theta\), with \(\beta = 7.0\) and \(\tilde{u} = 0.2\). The torque results from both the natural Brownian gyration and the applied shear load. To operate as a heat engine, the system must exhibit a torque reversal while remaining stable, meaning the shear-induced torque counteracts the natural Brownian gyration. The black dashed lines indicate regions where the system becomes unstable before reaching the stalling condition (zero torque). The green curve represents a scenario where the system remains stable but does not function as either a heat engine or a refrigerator. The inset provides a log-scale view, emphasizing the torque reversal where \(\langle \tau \rangle\) changes sign.
	}
	\label{figure04}
\end{figure}
In Fig.~\ref{figure04}, we represent the average torque \(\langle \tau \rangle\) as a function of the dimensionless shear rate \(\dot{\tilde{\gamma}}\) for different shear orientations \(\theta\). The sign change of \(\langle \tau \rangle\) indicates the transition from a regime where the natural gyration dominates to one where the shear force overcomes the temperature-driven rotation. For the system to function as a heat engine, the torque generated by the shear must act against the natural gyration caused by the temperature asymmetry. The black dashed lines correspond to unstable parameter regions where the steady-state operation cannot be maintained. Interestingly, in some cases, the system reaches instability before the stalling condition (\(\langle \tau \rangle = 0\)), implying that while certain shear orientations enable efficient energy extraction, they also lead to destabilization at intermediate shear rates. The green curve represents a scenario where the system remains stable but does not function as an engine or refrigerator.


\begin{figure*}[t]
	\centering
	\includegraphics[width=\linewidth]{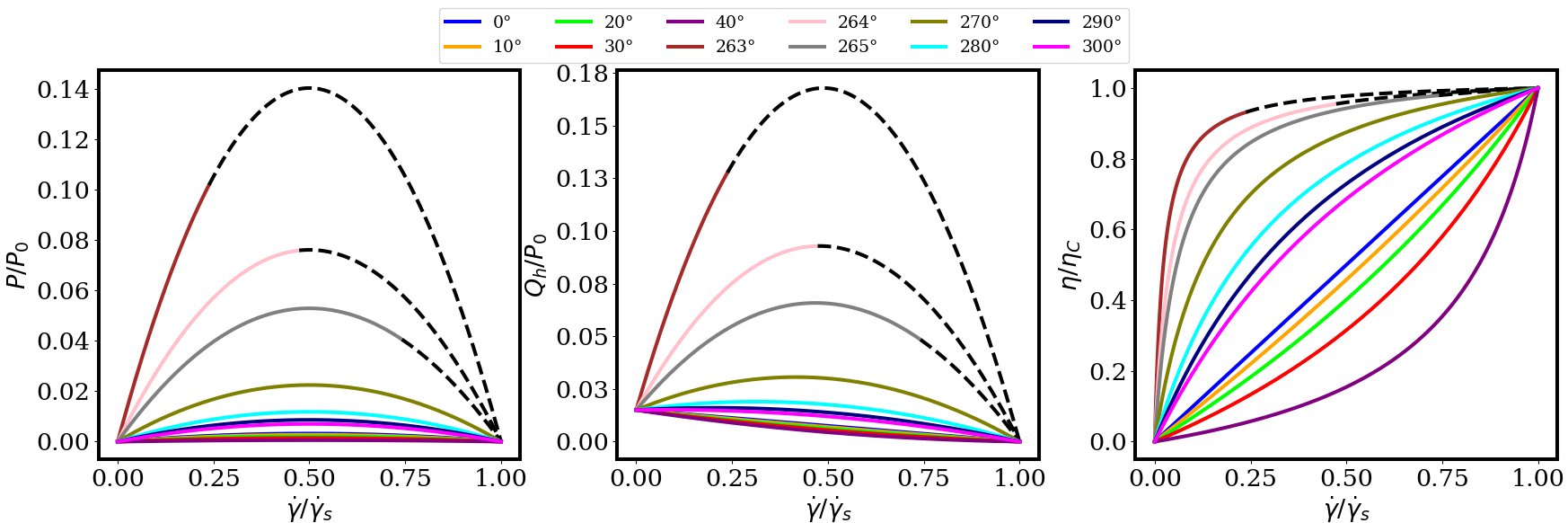}
	\caption{Panel showing, from left to right, the average mechanical power \(P/P_0\), the extracted heat from the hot heat bath \(Q_h/P_0\), and the scaled efficiency \(\eta/\eta_C\) as functions of the shear rate \(\dot{\gamma}/\dot{\gamma}_s\) for different shear directions \(\theta\). The parameters used are \(\beta = 7.0\) and \(\tilde{u} = 0.2\), with \(P_0 = k(T_c + T_h)/\gamma\) serving as the characteristic power scale. The black dashed lines indicate regions where the system becomes unstable before reaching the stalling condition. The Carnot efficiency is theoretically attainable at the stalling point where no mechanical power can be extracted. Remarkably, for certain shear directions, particularly those corresponding to flow orientations that enhance the conversion of thermal energy into work, the efficiency approaches Carnot values even at finite power output. This highlights the nontrivial role of shear-induced loading in optimizing micro-scale heat engines.
	}
	\label{figure05}
\end{figure*}

\subsection{Efficiency}

To evaluate the performance of the Brownian gyrator as a heat engine, we quantify its efficiency in converting thermal energy into mechanical work against the applied shear force. The efficiency, defined as the ratio of extracted mechanical power to the heat absorbed from the hot reservoir, provides insight into the engine’s effectiveness in utilizing the temperature gradient for energy conversion. 

In stochastic thermodynamics, energy transfer occurs continuously through small fluctuations, unlike macroscopic heat engines, which operate in cyclic steps. Here, the heat absorbed from the hot bath contributes to both mechanical work and dissipation into the cold bath, following the fundamental relation $\langle \dot{Q}_h \rangle = \langle \dot{W} \rangle + \langle \dot{Q}_c \rangle$. By analyzing these quantities in the steady-state regime, we determine how the efficiency varies with the shear rate and orientation. 

Following the rules of stochastic thermodynamics, along an individual trajectory we identify the work

\begin{equation}
	\label{work_def}
\dbar{W} = -d\mathbf{r} \cdot \mathbf{F}^{\text{shear}}(\mathbf{r}),
\end{equation}
performed against the external force. Using the above equation we can obtain the average mechanical power $\langle\dot{W}\rangle=\langle\dbar{W}/dt\rangle = -\dot{\gamma}[\langle xv_y\rangle\sin(\theta)+\langle yv_x\rangle\cos(\theta)]$, which reads

\begin{equation}
	\label{work}\small{
	\langle\dot{W}\rangle\! =\! \frac{\dot{\gamma}k}{2\gamma } (\cos(\theta)\!-\!\sin(\theta))\left[\tilde{u}(T_h\!-\!T_c)\!-\!\dot{\tilde{\gamma}}(T_h\cos(\theta)\!+\! T_c\sin(\theta))\right], }
\end{equation}
with the stall parameter

\begin{equation}
	\label{stall_parameter}
	\dot{\tilde{\gamma}}_s = \frac{u\eta_C}{\cos(\theta)-(1-\eta_C)\sin(\theta)},
\end{equation}
which is the point at which the gyration is fully suppressed and no work can be extracted. Beyond the stalling shear rate the system ceases to function as a heat engine.

The rate of the heat extracted from the hot heat bath can be calculated as

\begin{equation}
	\label{heath_def}
	\dbar{Q_h} = -dy[F_y^c + F_y^{shear}],
\end{equation}
and the rate of the heat dissipated into the cold heat bath as

\begin{equation}
	\label{heatc_def}
	\dbar{Q_c} = dx[F_x^c + F_x^{shear}].
\end{equation}
The total heat flux is the rate of work done by the engine whose average is given by $\langle\dot{W}\rangle = \langle\dot{Q}_h\rangle -\langle\dot{Q}_c\rangle $. The heat out of the hot heat bath is needed for the calculation of the efficiency, which can be calculated and written as

\begin{equation}
	\label{heath}\small{
	\langle\dot{Q}_h\rangle\! =\! \frac{k(\tilde{u}\!-\!\dot{\tilde{\gamma}}\sin(\theta))}{2\gamma } \left[\tilde{u}(T_h\!-\! T_c)\!-\!\dot{\tilde{\gamma}}(T_h\cos(\theta)\!+\! T_c\sin(\theta))\right],}
\end{equation}
and the efficiency

\begin{equation}
	\label{efficincy}
	\eta = \frac{\langle\dot{W}\rangle}{\langle\dot{Q}_h\rangle} = \frac{\dot{\tilde{\gamma}}(\cos(\theta)-\sin(\theta))}{\tilde{u}-\dot{\tilde{\gamma}}\sin(\theta)}
\end{equation}

In Fig.~\ref{figure05}, we analyze the performance of the heat engine by presenting the extracted mechanical power, the heat absorbed from the hot reservoir, and the corresponding efficiency, all as functions of the normalized shear rate $\dot{\tilde{\gamma}}/\dot{\tilde{\gamma}}_s$ for different shear orientations $\theta$. The leftmost panel depicts the mechanical power output, showing that for certain angles, power extraction reaches a peak before gradually diminishing as the shear rate approaches the stalling condition. The middle panel illustrates the heat absorbed from the hot bath, which follows a similar trend, highlighting the role of thermal fluctuations in sustaining the engine's operation. 

The rightmost panel of Fig.~\ref{figure05} shows the efficiency normalized by the Carnot efficiency, $\eta/\eta_C$, as a function of the shear rate. Notably, the system attains remarkably high efficiencies for specific shear directions, approaching the Carnot limit at stalling. This result underscores the significant role of shear orientation in optimizing energy conversion. However, the black dashed curves indicate regions where the system becomes unstable before reaching the stalling point, imposing a practical constraint on engine operation. The findings demonstrate that while high efficiency is achievable, it is often accompanied by an inherent stability trade-off.

\begin{figure}[t]
	\centering
	\includegraphics[width=8cm]{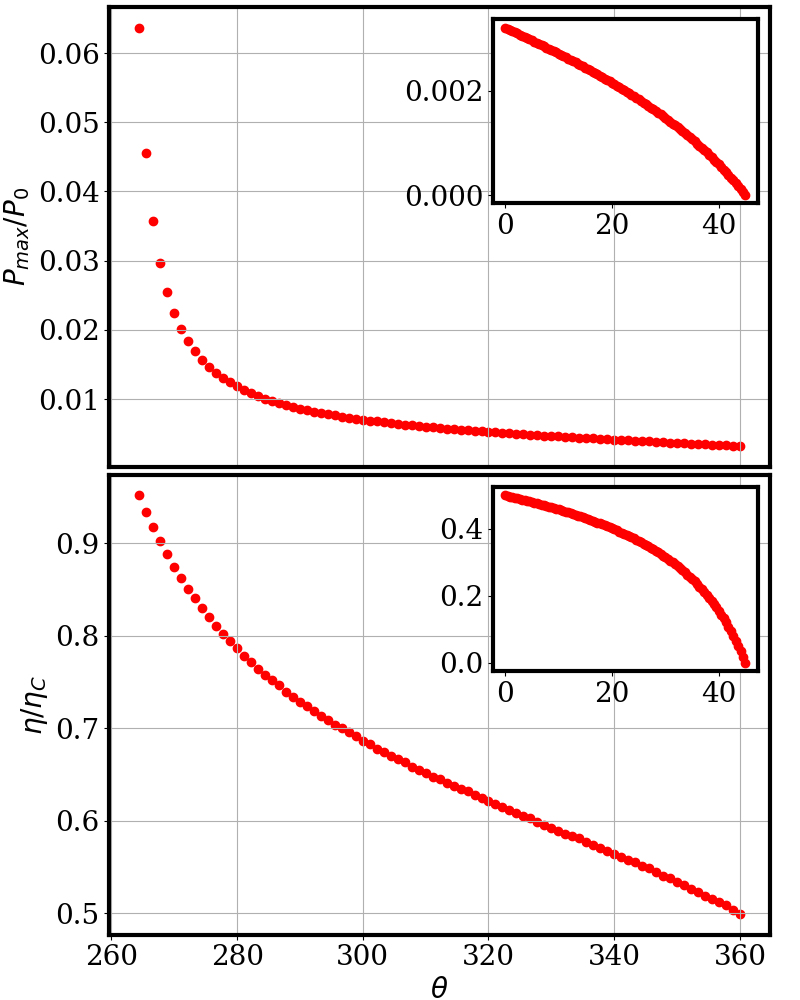}
	\caption{    The maximum scaled power \( P_{\max}/P_0 \) (top panel) and the corresponding scaled efficiency at maximum power \( \eta_{\max}/\eta_C \) (bottom panel) as functions of the shear direction \(\theta\), extracted from Fig.~\ref{figure05}. These values are shown only for shear directions where the system remains stable at maximum power. The insets highlight the behavior for small \(\theta\), demonstrating the rapid decrease in power and efficiency as \(\theta\) deviates from optimal directions. The results indicate that efficiency can approach Carnot efficiency at maximum power, which depends sensitively on the shear direction relative to the temperature axes, reinforcing the crucial role of directional shear in optimizing heat-to-work conversion. The parameters used are \(\beta = 7.0\) and \(\tilde{u} = 0.2\), with \(P_0 = k(T_c + T_h)/\gamma\) serving as the characteristic power scale.  
	}
	\label{figure06}
\end{figure}

Figure~\ref{figure06} summarizes the maximum extracted power (top panel) and the corresponding efficiency at maximum power (bottom panel) as functions of the shear orientation $\theta$. The results reveal that maximum power extraction is strongly dependent on the shear direction, with the highest values occurring near angles where the natural gyration is effectively opposed by the applied shear force. The lower panel highlights that for certain orientations, the efficiency at maximum power remains remarkably high, reaching a substantial fraction of the Carnot efficiency. These findings suggest that by carefully tuning the shear direction, one can achieve an optimal balance between power output and efficiency, although stability constraints must always be considered.


\section{Discussion}
Our study unveils the intricate interplay between Brownian gyration and externally applied shear flow, demonstrating how this coupling can drive a microscopic heat engine with remarkable performance. By analytically deriving the steady-state probability distribution, we quantified the average torque exerted on the potential, the mechanical power extracted from thermal fluctuations, and the heat dissipated to the surrounding baths. The results reveal an unexpected yet fundamental trade-off between efficiency, power output, and stability, offering new insights into the limits of small-scale heat engines operating under external forcing.

A striking finding of our analysis is that for specific shear orientations, the system can approach Carnot efficiency at maximum power—an extraordinary outcome in finite-time thermodynamics. This enhancement is driven by an optimized alignment of the shear load with the natural gyration, allowing the system to efficiently harness energy from the temperature gradient. However, this extreme efficiency comes at a cost: in certain parameter regimes, the system becomes unstable before reaching the stall condition. This instability manifests as increased fluctuations in position, ultimately disrupting the steady-state operation of the engine. Such a trade-off underscores the delicate balance between dissipation and control in non-equilibrium systems, where achieving optimal performance may require operating near the boundary of stability.

Moreover, our findings reveal that shear flow not only counteracts the natural gyration but also fundamentally alters the system's phase-space dynamics. In some cases, the shear introduces an effective stabilizing force, reinforcing confinement and allowing the system to sustain its cyclic motion even at high shear rates. In other cases, it acts as a destabilizing force, increasing fluctuations and driving the system out of its stable steady state. This dual role of shear is essential in explaining why some shear orientations allow highly efficient engine operation while others lead to instability before reaching the stall condition. Such findings may have broader implications in the design of nanoscale engines, suggesting that external driving forces can be tailored not only to optimize performance but also to regulate stability.

The insights gained from this study extend beyond the Brownian gyrator and may have relevance for a wide range of non-equilibrium systems, including active matter, colloidal transport, and microscale engines. The observed trade-off between efficiency and stability is reminiscent of constraints found in biological systems, where fluctuating forces must be carefully controlled to maintain functionality. Future experimental realizations of shear-driven Brownian engines, using optical tweezers or microfluidic setups, could provide direct verification of our theoretical predictions and offer a pathway for engineering microscopic machines with tunable performance characteristics~\cite{ziehl2009direct, holzer2010dynamics, argun2017experimental, polimeno2018optical, PhysRevE.77.041107}. From future perspective, it would be interesting to study how shear affects the escape dynamics in our studied model~\cite{kienle2011shear}.

\begin{acknowledgements}
We sincerely thank Kristian Stølevik Olsen for fruitful discussions that provided valuable insights into this work. We gratefully acknowledge funding from the Deutsche Forschungsgemeinschaft (DFG) through the SPP 2265 under Grant No. LO 418/25-2 (I.A. and H.L.) and support for A.S. under Project No. SH 1275/5-1.	
\end{acknowledgements}

\section*{AUTHOR DECLARATIONS}
\subsection*{Conflict of Interest}
The authors have no conflicts to disclose.

\section*{DATA AVAILABILITY}
The data that support the findings of this study are available
from the corresponding author upon reasonable request.
\appendix



\section{Steady-state solution} \label{appendixA}
In this appendix, we provide the detailed derivation of the steady-state solution of the Fokker-Planck equation governing the dynamics of the system. We explicitly obtain the steady-state probability density function (PDF), ensuring proper normalization, which can be used to derive the corresponding probability fluxes, characterizing the persistent rotational motion in the system. 

We recall the linear multivariate Fokker–Planck equation, given in Eq.~\eqref{FPE_time}

\begin{equation}
	\label{FPE_app}
	\frac{\partial \rho(\mathbf{r}, t)}{\partial t} = -\nabla\cdot\left[\hat{\mathbf{A}}\mathbf{r} -  \hat{\mathbf{D}}\nabla \right]\rho(\mathbf{r}, t),
\end{equation}
with the diffusion matrix $\hat{\mathbf{D}}=\frac{1}{\gamma}\mathrm{diag}(T_x, T_y)$ and the drift matrix $\hat{\mathbf{A}}$ as

\begin{equation}
	\label{driftmatrix_app}
	\hat{\mathbf{A}} = -\frac{k}{\gamma}\begin{pmatrix}
		1 & \tilde{u}-\dot{\tilde{\gamma}}\cos(\theta) \\ \tilde{u}-\dot{\tilde{\gamma}}\sin(\theta) & 1 
	\end{pmatrix}.
\end{equation}
This equation has a Gaussian solution which can be written as 
\begin{equation}
	\label{solution_time_app}
	\rho(\mathbf{r}, t) = \frac{1}{Z} e^{-\mathbf{r}^\top\cdot\hat{\mathbf{C}}_t^{-1}\cdot\mathbf{r}},
\end{equation}
where \(Z=2\pi\sqrt{\mathrm{Det}(\hat{\mathbf{C}}_t)}\) is the normalization factor. the covariance matrix \(\hat{\mathbf{C}}_t\) satisfies the following Lyapunov equation
\begin{equation}
	\label{Lyapunov_app}
	\frac{d\hat{\mathbf{C}}_t}{d t} = \hat{\mathbf{A}}\hat{\mathbf{C}}_t + \hat{\mathbf{C}}_t \hat{\mathbf{A}}^\top + 2 \hat{\mathbf{D}},
\end{equation}
which can be solved for the stationary state by
setting the left-hand side to zero. This gives the stationary-state covariance matrix, denoted by \(\hat{\mathbf{C}}\), whose elements $C_{ij}$ with $i, j \in \{x, y\}$ are given as
{\small{
\begin{align} 
	C_{xx} & \!=\! \frac{1}{2k}\left[T_c\!+\!\frac{T_c\!+\!\tilde{u}^2T_h\!+\!\dot{\tilde{\gamma}}T_h\cos(\theta)\left(-2\tilde{u}\!+\!\dot{\tilde{\gamma}}\cos(\theta)\right)}{1\!-\!\tilde{u}^2\!+\!\tilde{u}\dot{\tilde{\gamma}}\sin(\theta)\!+\!\dot{\tilde{\gamma}}\cos(\theta)\left(\tilde{u}\!-\!\dot{\tilde{\gamma}}\sin(\theta)\right)}\right], \\
	C_{xy} & \!=\! -\frac{1}{2k}\left[\frac{\tilde{u}(T_c+T_h)-\dot{\tilde{\gamma}}\left(T_h\cos(\theta)-T_c\sin(\theta)\right)}{1\!-\!\tilde{u}^2\!+\!\tilde{u}\dot{\tilde{\gamma}}\sin(\theta)\!+\!\dot{\tilde{\gamma}}\cos(\theta)\left(\tilde{u}\!-\!\dot{\tilde{\gamma}}\sin(\theta)\right)}\right], \\
	C_{yx} & =C_{xy}, \\
	C_{yy} & \!=\! \frac{1}{2k}\left[T_h\!+\!\frac{T_h\!+\!\tilde{u}^2T_c\!+\!\dot{\tilde{\gamma}}T_c\sin(\theta)\left(-2\tilde{u}\!+\!\dot{\tilde{\gamma}}\sin(\theta)\right)}{1\!-\!\tilde{u}^2\!+\!\tilde{u}\dot{\tilde{\gamma}}\sin(\theta)\!+\!\dot{\tilde{\gamma}}\cos(\theta)\left(\tilde{u}\!-\!\dot{\tilde{\gamma}}\sin(\theta)\right)}\right], 
\end{align}
}with $Z$ given in the main text. By calculating the inverse of the steady-state covariance matrix we obtain the steady-state probability density given in Eqs.~\eqref{pdf_ss}-\eqref{normalization} in the main text, which can be used to compute the probability fluxes as in Eq.~\eqref{fluxes}. 

\section{Stability Condition}

The stability of the system can be determined by the eigenvalues of the drift matrix $\hat{\mathbf{A}}$. For the system to maintain a steady-state gyration, all eigenvalues must have negative real parts, ensuring that fluctuations decay over time. The eigenvalues of $\hat{\mathbf{A}}$ are given by

\begin{equation}
	\lambda_{\pm}=\frac{k}{\gamma} \left[-1\pm \sqrt{(\tilde{u}-\dot{\tilde{\gamma}}  \sin (\theta )) (\tilde{u}-\dot{\tilde{\gamma}}  \cos (\theta ))}\right].
\end{equation}

Since the first term in the expression for $\lambda_{\pm}$ is always negative, the system remains stable under the following conditions:  
(i) If the term under the square root is negative, the eigenvalues remain complex with a negative real part, ensuring stable oscillatory behavior.  
(ii) If the term under the square root is positive, both eigenvalues are real, and stability is maintained only if both eigenvalues remain negative. This requires that the square root term does not exceed unity, i.e.,

\begin{equation}
	\sqrt{(\tilde{u}-\dot{\tilde{\gamma}}  \sin (\theta )) (\tilde{u}-\dot{\tilde{\gamma}}  \cos (\theta ))} < 1.
\end{equation}

This analysis, which has been taken into account throughout the paper, establishes a clear stability criterion for the operational regime of the Brownian gyrator under shear, imposing the limits within which it can function as a microscopic heat engine.

\section*{References}


%
\end{document}